\documentclass[conference]{IEEEtran}
\usepackage{graphics}
\usepackage{graphicx}
\usepackage{epsfig,amssymb}
\usepackage{epstopdf}
\usepackage{amsmath}
\usepackage{breqn}
\usepackage{times}
\usepackage{mathrsfs}
\usepackage{array}
\usepackage{amsmath, latexsym, amsfonts, amssymb}
\usepackage[mathscr]{eucal}
\usepackage{array, tabularx}
\usepackage[table]{xcolor}
\usepackage{multirow}
\usepackage{epsfig}

\usepackage{cite}
\usepackage[numbers,sort & compress]{natbib}

\usepackage{tikz}
\usepackage{url}
\usepackage{mathtools}
\usepackage{psfragx}
\usepackage{xcolor}
\usepackage{bbm}
\usepackage{authblk}
\newcommand{\paren}[1]{\left(#1\right)}
\newcommand{\sqparen}[1]{\left[#1\right]}
\newcommand{\brparen}[1]{\left\{#1\right\}}

\newcommand{\R}{\ensuremath{\field{R}}} 
\newcommand{\C}{\ensuremath{\field{C}}} 
\newcommand{\defeq}{\ensuremath{\triangleq}} 
\renewcommand{\vec}[1]{\ensuremath{\boldsymbol{#1}}} 
\newcommand{\ie}{\ensuremath{{\text{\em i.e.}}}}

\newcommand{\D}{\displaystyle}

\newtheorem{theorem}{Theorem}
\newtheorem{lemma}{Lemma}

\IEEEoverridecommandlockouts

\def\B{
	\begin{bmatrix}
		1 \\
	e^{j\theta_1} \\
		0
	\end{bmatrix}}
	
\def\C{
	\begin{bmatrix}
		1 \\
	e^{j\theta_2} \\
	0
	\end{bmatrix}}
		
\def\D{
	\begin{bmatrix}
		1 \\
	e^{j\theta_3} \\
		0
	\end{bmatrix}}
	
\def\E{
	\begin{bmatrix}
		1 \\
		0 \\
		0
	\end{bmatrix}}

	\def\P{
	\begin{bmatrix}
		1 \\
	0\\
		e^{j\theta_1} 
	\end{bmatrix}}
	
\def\Q{
	\begin{bmatrix}
		1 \\
		0\\
		e^{j\theta_2} 
\end{bmatrix}}
		
\def\R{
	\begin{bmatrix}
		1 \\
		0\\
		e^{j\theta_3} 
	\end{bmatrix}}
	
\def\S{
	\begin{bmatrix}
		1 \\
	0 \\
		0
	\end{bmatrix}}		
	
\begin{document}

\title{Opportunistic  Wireless Energy Beamforming using RSSI Feedback}

\author[$\ast \dag $]{Samith Abeywickrama}
\affil[$ \ast $]{National University of Singapore, Singapore}
\affil[$ \dag $]{Singapore University of Technology and Design,  Singapore}


\affil[$  $ ]{Email: \textit { samith@u.nus.edu}}

\date{}
\bibliographystyle{ieeetr}
\maketitle

\begin{abstract}
This paper proposes a novel  channel
estimation method, and a cluster-based opportunistic scheduling
policy, that enhances the efficiency of wireless energy transfer
in a system consisting of multiple energy receivers (ERs).
Firstly, in the training stage, the energy transmitter (ET) obtains a set of RSSI feedback values from all ERs, and these values are used to estimate the channels between the ET and all ERs.
Next, based on the channel estimates, the ERs are grouped into clusters, and the cluster that
has its members closest to its centroid in phase is picked. This cluster will be given priority, and the beamformer that maximizes the minimum harvested energy among all ERs in the selected cluster is found by solving a convex optimization problem. This beamformer is used to transfer energy to the ERs using maximum ratio transmit beamforming.  All
ERs have the same chance of being selected regardless of their
distances from the ET, and hence, this scheduling policy can be
considered to be opportunistic as well as fair. It is shown that the proposed method achieves significant performance gains. 
\end{abstract}

\section{Introduction}


Radio frequency (RF) signal enabled wireless energy transfer (WET) using multiple antennas at the energy transmitter (ET), has become a promising technology for facilitating convenient and perpetual power supply to charge freely located devices \cite{power1}. Increasing the efficiency of the energy transfer between the ET and the energy receiver (ER) is of paramount importance for this process. The availability of channel state information (CSI), and the scheduling policy when multiple ERs are present, can be identified as two main contributors to this efficiency. To this end, this paper proposes a novel  channel estimation method, and a cluster-based opportunistic scheduling policy, that enhances the efficiency of a WET system consisting of multiple ERs.

WET performed utilizing CSI normally consists of a training stage for channel learning. However, due to tight energy constraints and hardware limitations at the ERs, the conventional pilot-based techniques, where the channel estimation is done at the ERs, lead to many implementation difficulties \cite{8362859,8362863}. The authors of \cite{one_bit} sought to estimate the channel at the ET using a one-bit feedback algorithm, where phase perturbations are made based on the feedback bits to obtain a satisfactory beamforming vector for WET. \cite{recip3} proposes exploiting the channel reciprocity for channel learning, which, however, is practically difficult to realize \cite{8254920}. \cite{rssi_work, our_rssi, our_tsp} propose energy efficient channel estimation methods based on RSSI values that are fed back from the receiver to the transmitter, and out of them, \cite{our_rssi,our_tsp} can be considered to be the most related to our work. Moreover, \cite{our_rssi,our_tsp}  propose a methodology of estimating the phase values of the channels between a single ER and each antenna of the ET, and these estimates are utilized to employ Equal Gain Transmit (EGT) beamforming for WET. In this paper, we consider multiple ERs, and focus on utilizing RSSI values to estimate both channel phase and channel magnitude information using a maximum likelihood analysis, in order to perform more superior Maximum Ratio Transmit (MRT) beamforming. 

Although we have enough information to employ MRT beamforming after the channel estimation, due to having multiple ERs, we end up with an interesting scheduling problem that will definitely affect the effectiveness of the WET. To this end, \cite{oppor1} proposes an opportunistic scheduling policy for WET, where the beamformer is designed based only on the ER having the best channel. This method achieves improved performance compared to the conventional round-robin scheduling. In \cite{oppor2}, random beamforming is employed, where ET randomly designs a beamformer regardless of the channel information of ERs. This scheme ensures fairness. Our proposed scheduling policy is both opportunistic and fair, and is as follows. Based on the channel estimates between the ET and the ERs, we group the ERs into clusters using the Lloyd's Algorithm \cite{lloyd}, and we pick the cluster that has its members (ERs) closest to its centroid. All the ERs in the network will harvest energy in the wireless power beamforming (WPB) stage. However, we will give priority to the selected cluster. To this end, we solve a convex optimization problem to find the beamformer that maximizes the minimum harvested energy among all ERs in the selected cluster. In this optimization problem, focusing on a set of ERs that are close to each other in channel phase leads to a significant improvement in the design of the beamformer, and since all ERs have the same chance of being in the selected cluster  regardless of their
distances from the ET, the scheduling policy is fair over time as well.

The paper organization is as follows. The system model and the problem setup is in Section \ref{Section:System model}. Section \ref{tra} discusses the channel estimation, and Section \ref{opt} discusses how the optimization problem can be solved. Then, in Section \ref{results}, we highlight significant performance improvements that can be obtained thanks to the proposed estimation method and the scheduling policy, through simulations. Section \ref{conclusions} concludes the paper.




\section{System Model and Problem Setup}\label{Section:System model}

We consider a MISO channel for WET. An ET consisting of $K \geq 2$ antennas delivers energy over a wireless medium to $N$ ERs, each equipped with a single receive antenna. 
The ET transmits $M$ beams along the direction of $M$ beamforming vectors $\brparen{\vec{b_m} \in \mathbb{C}^{K \times 1}}^M_{m=1}$, such that the transmit signal at the ET is given by $$ \vec{x} = \sum_{m=1}^M\vec{b}_m s, $$ where $ s $ denotes the transmit symbol, which is independent of the beamforming vectors, and $ \mathbb{E}( |s|^{2}) =1 $.  
It is assumed that the maximum transmit sum-power constraint at the ET is $ P > 0 $.  Therefore, we have $ \mathbb{E} (\|\vec{x}\|^{2}) = \mathrm{tr}(\mathrm {\bf C}_{\vec {xx}}) \leq P$, where $  \mathrm {\bf C}_{\vec {xx}} = \mathbb E (\vec{x}\vec{x}^{\dag})$ is the transmit covariance  matrix, and $\mathrm{tr}(\cdot)$ and $ \|\cdot \| $ denote the trace of a square matrix and the Euclidean norm, respectively.

Let $ \vec{h}_i  \in \mathbb{C}^{K \times 1}$ represent the random complex MISO channel vector between the ET and the $i$-th ER, such that $\vec{h}_i = \sqparen {|h_{i,1}|e^{j\delta_{i,1}}, \dots , |h_{i,K}|e^{j\delta_{i,K}}}^\top $. Moreover, for the simplicity in notations, $\vec{h}_i$ is assumed to be the product of the path loss and multipath fading  between the ET and the $i$-th ER, and it is considered to be independent and identically distributed (i.i.d.) with an arbitrary distribution. The received energy (or RSSI) at the $i$-th ER can be written as
\begin{eqnarray}
\mathrm R_i = \xi (\vec h^{\dag}_i \mathrm {\bf C}_{\vec {xx}}\vec h_i), \label{power}
\end{eqnarray}
where $\xi$ denotes the conversion efficiency of the ER \cite{power1}. We assume $ \xi=1 $ for  simplicity, 
and consider a quasi-static block-fading channel model and a block-based energy transmission, where it is assumed that $\vec{h}_i$ remains constant over each transmission block.


It is well known that CSI is vital for the efficiency of beamforming. Therefore, the WET process consists of two stages. Firstly, we have the training stage that the ET uses for channel learning. Then, the knowledge on the channel is used to set the beamforming vectors for the second stage, that we call the wireless power beamforming (WPB) stage, where the actual WET happens. 
Since we are particularly focusing on applications having tight energy constraints at the ERs, performing channel estimation at the ER may become infeasible as it involves analog to digital conversion and baseline processing, which require significant energy. We will obtain estimates of $\brparen{\vec{h}_i}_{i=1}^{N}$, by only considering RSSI values that are fed back from the ERs to the ET, and we will utilize these estimates to perform multi-user maximum ratio transmit (MRT) beamforming in the WPB stage.


After estimating $\brparen{\vec{h}_i}_{i=1}^{N}$, we cluster the ERs in to $Q$ clusters using the Lloyd's Algorithm \cite{lloyd}. Let $\mathcal{Q}^\star$ be the cluster that has its cluster members (ERs) closest to its centroid in channel phase, $\ie$, the cluster in which the sum of Euclidean distances between the ERs and the centroid of the cluster is minimum. It should be noted that ERs in the same cluster may or may not be close to each other spatially. All the ERs in the network will harvest energy in the WPB stage. However, we will give priority to the ERs in $\mathcal{Q}^\star$. 

The clustering is done by only considering the phase values of the estimated channel vectors due to the following  reasons. Firstly, since the phase values change rapidly over time (i.i.d. in our model), all ERs have the same chance of being in the selected cluster, which ensures fairness for the whole network over time. If the magnitudes of $\brparen{\vec{h}_i}_{i=1}^{N}$ are considered, the location dependent path loss values of the ERs, which  change slowly over time, will play a significant role  in  clustering, and thus, will affect the fairness in scheduling. Secondly, due to the phase values being uniformly and identically distributed, the cluster sizes will not differ significantly from each other. Note that, the sum of Euclidean distances between the ERs and a centroid of the cluster depends on the number of ERs in the cluster. Therefore, if there is a large variation in cluster sizes, $\mathcal{Q}^\star$ may end up being the smallest cluster with the lowest number of ERs, and this will not serve our purpose as well. We should stress that the notion of fairness in this paper is providing each ER in the network equal opportunity for being in $S^\star$ regardless of its distance from the ET, and being prioritized in the WPB stage. The harvested energy will differ among ERs depending on their distances from the ET.

We formulate an optimization problem to design a beamformer that maximizes the minimum harvested energy among all ERs in $\mathcal{Q}^\star$. This will ensure fairness for all ERs in $\mathcal{Q}^\star$, and since the channel vectors are i.i.d., all ERs have the same chance of being in the selected cluster. Therefore, the scheduling policy will ensure fairness for all ERs in the network over time. 
Let $ \hat {\vec{h}}_i  $ denote the estimated channel vector from the training stage for $i \in \mathcal{Q}^\star$, and let $ \vec \eta_i $ denote the channel estimation error.  $ \vec \eta_i $ is assumed to be bounded, \textit{i.e.}, $ \|\vec \eta_i \|_{\mathrm F} = \sqrt{\vec \eta_i^{\dag} \vec \eta_i }  \leq \varepsilon_i $, where $  \| \cdot \|_{\mathrm F} $ denotes the Frobenius norm and $ \varepsilon_i \geq 0 $.\footnote{ It should be noted that the estimation error is actually unbounded since we have considered Gaussian noise, and a probabilistic constraint may have  been more suitable. We have assumed bounded channel estimation uncertainties for the analytical tractability of the problem.  Please refer to \cite{bounded} where a similar approximation is made, and the necessity and the fairness of the approximation are justified.} By using these notations, the received energy (or RSSI) at the $i$-th ER in $\mathcal{Q}^\star$ can be written as $ \xi (\hat {\vec h}_i + \vec \eta_i)^{\dag} \mathrm {\bf C}_{\vec {xx}} (\hat {\vec h}_i + \vec \eta_i) $. Thus, our optimization problem can be formulated as 
\begin{equation}
\begin{aligned}
& \underset{\footnotesize \mathrm {\bf C}_{\vec {xx}} \succeq 0,\ t \geq 0  }{\text{maximize}}
& & t  \\
& \text{subject to}
& & \hspace{-0.25cm}\mathrm C1: \hspace{-0.25cm} \underset{\footnotesize \|\vec \eta_i \|  \leq \varepsilon_i }{\text{min}}  (\hat {\vec h}_i + \vec \eta_i)^{\dag} \mathrm {\bf C}_{\vec {xx}} (\hat {\vec h}_i + \vec \eta_i) \geq t \ \ \forall i \in \mathcal{Q}^\star \\
& &&  \hspace{-0.25cm} \mathrm C2: \mathrm{tr}(\mathrm {\bf C}_{\vec {xx}}) \leq P, 
\end{aligned}\label{Optimization Problem-MU} 
\end{equation} 
where $ t $ is a real-valued optimization variable. The problem is convex, but it is complex due to $ \mathrm C1 $ having infinitely many inequalities. Also, it has been shown in \cite{rank_one}, that if an optimization problem of the form in \eqref{Optimization Problem-MU} is solvable, the rank of the solution is one, $\ie$, optimality is achieved when $\mathrm{rank}(\mathrm {\bf C}_{\vec {xx}})=1$. This means, it is optimal to transmit a single beam on the downlink for WET. This sheds further light into why clustering will be useful in this context, as we will be better off focusing on a set of ERs that are closer to each other than considering all ERs, when designing the beamforming vector for this beam.  

In Section \ref{tra}, we will first discuss how the estimates of $\brparen{ {\vec{h}}_i}_{i=1}^{N}$ can be obtained by using RSSI feedback values. Then we will use these estimated values to cluster the ERs. Then, Section \ref{opt} will discuss how the optimization problem of interest can be solved by designing the beamformer that maximizes the minimum harvested energy among all the ERs in $\mathcal{Q}^\star$.

\section{Training Stage and channel estimation} \label{tra}

In \cite{our_rssi}, $K$ and $N$ are assumed to be 2 and 1, respectively, and a method of utilizing RSSI feedback values to estimate the phase difference of the two MISO channels between the ER and the ET has been proposed. This method can be directly used to estimate $\brparen{\phi_{i,v}}_{v=2}^K$, where $ \phi_{i,v} = \delta_{i,v} - \delta_{i,1} $. That is, for a given ER $i$, we can estimate the phase values of all channels between the ER and ET, relative to the channel between the ER and the first antenna of the ET (the first antenna is selected as reference without any loss of generality). Therefore, with the training stage proposed in \cite{our_rssi}, and \cite{our_tsp}, where an extension for $K>2$ is proposed, the ET can only employ equal gain transmit (EGT) beamforming in the WPB stage. With EGT beamforming, the ET equally splits the power among the transmit antennas, and pre-compensates channel phase shifts such that the signals are coherently added up at the ER regardless of the channel magnitudes.
In this paper, we employ MRT beamforming during WET stage, therefore the ET has to estimate $\brparen{\phi_{i,v}}_{v=2}^K$ as well as the channel magnitudes $\brparen{|h_{i,k}|}_{k=1}^K$ of all $i \in \brparen{1, \ldots, N}$, and hence, we need a modified training stage to facilitate these improvements.

\begin{figure}[t] \vspace{0.2cm}
	\centering {\includegraphics[trim = 0mm 0mm 0mm 0mm, clip,scale=0.7]{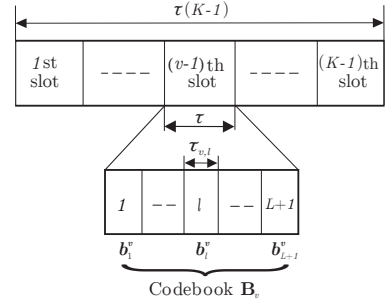}} 
	\caption{Training Stage.}	 
	\label{2pha} 
\end{figure}

We will start by defining a set of codebooks. For $v \in \brparen{2, \ldots, K}$, we define codebook $ \mathrm {\bf {B}}_v = \sqparen{\vec{b}_{1}^v \ 	\ldots \ \vec{b}_{L+1}^v} $ that includes $(L+1) $ complex $K$-by-$1$ beamforming vectors. To this end, the $l$-th element of $\mathrm {\bf {B}}_v $ takes the form of $\vec{b}_{l}^v = \sqrt{\frac{P}{2}} \sqparen {b_{l,1}^v\ \   \cdots\ \  b_{l,K}^v}^\top $. Moreover, we have
\begin{alignat}{1}
b_{l,k}^v = \begin{cases}
1   & \text{if } k=1  \\
\exp\paren{j\theta_l}    & \text{if } k=v \ \ \text{and} \ \ l \neq L+1  \\
0  & \text{otherwise}  \\
\end{cases},
\end{alignat}
where $ \theta_l =  \frac{2(l-1)\pi}{L} $. The training stage consists of $K-1$ time slots as illustrated in Fig. \ref{2pha}. In time slot $v-1$, the ET sequentially transmits using the $L+1$ beamforming vectors in $\mathrm {\bf {B}}_v $ for all $v \in \brparen{2, \ldots, K}$. Each ER will measure the respective $(K-1) \times (L+1)$ RSSI values for this sequential transmission, and will feed them back to the ET over orthogonal feedback channels \cite{orthoganal_feedback}. Note that the time taken for the channel learning does not depend $ N $, and it depends only on $ K $ and $L$. 

For the clarity of understanding, we will explain the structure of the codebooks by using an example. Consider that $ K = 3 $ and $ L=3 $. For this selection, we have 
\begin{equation*}
\mathrm {\bf {B}}_2  = \sqrt{\frac{P}{2}} \left(\B \C \D \E \right)_{3 \times 4}, 
\end{equation*}
and
\begin{equation*}
\mathrm {\bf {B}}_3  = \sqrt{\frac{P}{2}} \left(\P  \Q \R \S \right)_{3 \times 4}. 
\end{equation*}
There are two time slots in the training stage, and the ET will transmit using $\mathrm {\bf {B}}_2$ and $\mathrm {\bf {B}}_3$, respectively. From this example,
it is not hard to see that for all beamforming vectors except the last one in each codebook, the ET employs a pairwise antenna activation policy. To be more general, for the first $L$ beamforming vectors transmitted in time slot $v-1$, the ET only activates the first antenna and the $v$-th transmit antenna, for all $v \in \brparen{2, \ldots, K}$.  

Next, we will provide further insights on the design of the codebooks by looking into the estimation process.
To this end, by using \eqref{power}, the RSSI at the $ i$-th ER for the $ l$-th ($ \leq L $) element (beam) of $\mathrm {\bf {B}}_v $ can be written as
\begin{alignat} {2}
\mathrm R_{i,l}^v
&=\alpha_{i,v} + \beta_{i,v} \cos\paren{\theta_l+\phi_{i,v}} + z_i,
\label{eq:ch rssi_2} 
\end{alignat} 
where $\alpha_{i,v} = \frac{\xi P}{4}(|h_{i,1}|^{2} + |h_{i,v}|^{2})$, $\beta_{i,v} = \frac{\xi P}{2}|h_{i,1}||h_{i,v}|$, and $\phi_{i,v}= \delta_{i,v} - \delta_{i,1}$. Although we have assumed a quasi-static block-fading channel, due to the effect of noise, the RSSI value will change from one measurement to the other. We have used random variable $ z_i $ to represent this effect. More specifically, $ z_i $ captures the effect of all noise related to the measurement process such as noise in the channel, circuit, antenna matching network and rectifier, and we assume the random variables to be i.i.d. additive Gaussian having zero mean and variance $ \sigma^2 $. Therefore, we assume that in a given transmission block, the randomness in \eqref{eq:ch rssi_2} is caused only by $ z_i $.

It can be seen from \eqref{eq:ch rssi_2} that $  \mathrm R_{i,l}^v $ depends on three unknown parameters $ \alpha_{i,v}$, $\beta_{i,v}$, and $ \phi_{i,v}$. Thus, the parameter vector for the estimation process can be written as $\vec \varphi = [\alpha_{i,v} \enspace \beta_{i,v} \enspace \phi_{i,v}]^{\top} $. For a given $v \in \brparen{2,\ldots,K}$, the ET will receive $L$ feedback values of the form of \eqref{eq:ch rssi_2} from each ER, and we will utilize these feedback values to estimate $\phi_{i,v}$ for each $i \in \brparen{1,\ldots,N}$, which gives us enough information to perform EGC. However, to perform MRT, we need amplitude information as well, thus, we have to estimate $ \alpha_{i,v}$ and/or $\beta_{i,v}$ as well. As discussed later, estimating either $ \alpha_{i,v}$ or $\beta_{i,v}$ is sufficient for our requirement, thus, we will estimate $ \alpha_{i,v}$. It should be highlighted, that we have selected $\brparen{\theta_l}_{l=1}^L$ in our codebooks in such a manner that the estimators of all three parameters of interest achieve the Cramer-Rao lower bound (CRLB). The CRLB is the best performance that an unbiased estimator can achieve as it gives a lower bound on the variance of an unbiased estimator \cite{8712420} \footnote{To be exact, the selection of $\brparen{\theta_l}_{l=1}^L$ minimizes the modified CRLB \cite{our_rssi}, and we skip the details due to space limitations}. Also, it can be shown by using the Fisher information matrix of $\vec \varphi$, that $L \geq 3$ for the estimation process to be possible \cite{our_tsp}.

Based on the assumption that the noise power is i.i.d. Gaussian, estimating $\phi_{i,v}$ and $ \alpha_{i,v}$ for a given $i$ and $v$ becomes a classical parameter estimation problem. A maximum likelihood estimate of these parameters can be obtained by finding the value of $\phi_{i,v}$ and $ \alpha_{i,v}$ that minimizes 
\begin{eqnarray}
\mathrm E \defeq \sum_{l=1}^{L} \sqparen{\mathrm R_{i,l}^v - \paren{\alpha_{i,v} + \beta_{i,v} \cos\paren{\theta_l+\phi_{i,v}}} }^{2}, \label{ml}
\end{eqnarray}  
respectively. These ideas are formally presented through the following theorem, and we have skipped the proof details due to space limitations.
\begin{theorem} \label{Thm:esti}
For the $i$-th ER, let $R_{i,l}^v$ denote the respective RSSI value for the $l$-th beamforming vector of $\mathrm {\bf {B}}_v $. 
Then $ \phi_{i,v} $  and $ \alpha_{i,v} $ can be estimated by   
\begin{gather}
\hat	 \phi_{i,v} = \tan^{-1}\paren{\frac{\displaystyle  -\sum_{l=1}^{L} \mathrm R_{i,l}^v \sin{\paren{\theta_l}} }           
	{\displaystyle \sum_{l=1}^{L} \mathrm R_{i,l}^v \cos{\paren{\theta_l}}}	
},
\label{estimation_phi} 
\end{gather}
and $ \hat{\alpha}_{i,v} = \displaystyle  \sum_{l=1}^{L} \mathrm R_{i,l}^v / L,  $
respectively, where $ \theta_l =  \frac{2(l-1)\pi}{L} $.
\end{theorem}
Note that the results in Theorem \ref{Thm:esti} are rather simple, very easy to calculate,
and requires minimal processing. We should stress that the manner in which we have selected $\brparen{\theta_l}_{l=1}^L$ have indirectly lead to the simplifications of these results as well. Ambiguity resolution in $\hat	 \phi_{i,v}$ can be done using similar techniques discussed in \cite{our_tsp}.

Next, let us focus on the $(L+1)$-th vector of each codebook. We have $\alpha_{i,v} = \frac{\xi P}{4}(|h_{i,1}|^{2} + |h_{i,v}|^{2})$, which we have already estimated. For MRT, we need $|h_{i,v}|$, and to extract this from $\alpha_{i,v}$, we need to know $|h_{i,1}|$. When $ l =L+1 $, we transmit using the first antenna only, and the corresponding RSSI value is given by 
\begin{alignat} {2}
\mathrm R_{i,(L+1)}^v =  \frac{\xi P}{2} |h_{i,1}|^{2} + z_i.
\label{eq:ch rssi_3} 
\end{alignat}
Estimating $|h_{i,1}|$ using the same concepts as earlier will allow us to recover estimates of $\brparen{|h_{i,k}|}_{k=1}^K$, which gives us enough information to perform WET using MRT, and also to cluster the ERs in to clusters using the Lloyd's Algorithm. We should note that if $K=2$, the ET will only receive one feedback value of the form in \eqref{eq:ch rssi_3}. For this case, we will have to repeat the $(L+1)$-th beamforming vector to get some more feedback values to facilitate the estimation process of $|h_{i,1}|$.


%
%



\section{Solution to the optimization problem} \label{opt}

The problem in \eqref{Optimization Problem-MU} is convex, but it is complex due to $ \mathrm C1 $ having infinitely many inequalities.
Therefore, this problem can be alleviated by transforming $ \mathrm C1 $ into a linear matrix inequality (LMI), 
and this is possible by applying the S-procedure. These ideas are formally presented through the following lemma. 
\begin{lemma} \label{Lemma:LMI}
	The equivalent LMI of constraint $ \mathrm C1 $ in (\ref{Optimization Problem-MU}) is given by $ \mathrm \mathrm {\bf T}_{i}(\mathrm {\bf C}_{\vec {xx}},t, \mu_i) \succeq 0 \ \ \forall i \in \mathcal{Q}^\star $,
	where
	\begin{equation*}
	\mathrm {\bf T}_{i}(\mathrm {\bf C}_{\vec {xx}},t, \mu_i) 
	= \\ 
	\begin{bmatrix}
	\mu_i  \mathrm {{{\bf{I}}}}_K+  \mathrm {\bf C}_{\vec {xx}}    &  \mathrm {\bf C}_{\vec {xx}} \vec h_i\\
	\vec h_i^{\dag} \mathrm {\bf C}_{\vec {xx}} &  \vec h_i^{\dag} \mathrm {\bf C}_{\vec {xx}}\vec h_i-t-\mu_i \varepsilon_i^2 \\
	\end{bmatrix} ,
	\end{equation*}
	and $ \mu_i \geq 0$ is a real-valued variable.
	\label{LMI}
\end{lemma}
\begin{IEEEproof}
	For $ g=1,2 $, let $ f_g(\vec \eta_i) $, be defined as 
	\begin{alignat} {2}
	f_g(\vec \eta_i)
	&= \vec \eta_i^{\dag} \mathrm{ \bf{A}}_g \vec \eta_i  + 2\text{Re} \{ \mathrm{\bf b}_g^{\dag} \vec \eta_i \} + c_g, \nonumber
	\end{alignat}
	where $\mathrm{ \bf{A}}_g \in \mathbb{C}^{K\times K}  $, $\mathrm{ \bf{b}}_g \in \mathbb{C}^{K\times 1}  $, and $ c_g \in \mathbb{R} $. The deduction (implication) $ f_1(\vec \eta_i) \leq 0 \Rightarrow f_2(\vec \eta_i) \leq 0 $ holds if and only if there exists $ \mu_i \geq 0$ such that 
	\begin{equation*}
	\begin{bmatrix}
	\mathrm{\bf A}_2   &  \mathrm{\bf b}_2\\
	\mathrm{\bf b}_2^{\dag} &  c_2 \\
	\end{bmatrix} 
	\preceq  \mu_i\\
	\begin{bmatrix}
	\mathrm{\bf A}_1   &  \mathrm{\bf b}_1\\
	\mathrm{\bf b}_1^{\dag} &  c_1 \\
	\end{bmatrix},
	\end{equation*}
	provided there exists a point $ \vec {\hat \eta_i} $ such that $ f_1(\vec {\hat \eta_i}) < 0 $. Now, with the focus of applying S-procedure, we write $ \mathrm {C1} $ as the
	following implication:
\begin{multline} 
{\vec \eta_i^{\dag} \mathrm {{{\bf{I}}}}_K \vec \eta_i }  \leq \varepsilon_i^2
\Rightarrow  -\vec \eta_i^{\dag} \mathrm {\bf C}_{\vec {xx}} \vec \eta_i  - \\ 2\text{Re} \{ \vec h_i^{\dag} \mathrm {\bf C}_{\vec {xx}} \vec h_i \vec \eta_i \}  - \vec h_i^{\dag} \mathrm {\bf C}_{\vec {xx}} \vec h_i + t\leq 0. \label{impli}
\end{multline}
	Using the definition of S-procedure on \eqref{impli}  
	completes the proof.
\end{IEEEproof}


Using this lemma, we can write the following equivalent optimization problem. 
\begin{equation}
\begin{aligned}
& \underset{\footnotesize \mathrm {\bf C}_{\vec {xx}} \succeq 0, t, \mu_i  }{\text{maximize}}
& & t  \\
& \text{subject to}
& & \mathrm C1:  \enspace \mathrm {\bf T}_{i}(\mathrm {\bf C}_{\vec {xx}},t, \mu_i) \succeq 0 \ \ \forall i \in \mathcal{Q}^\star\\
& &&  \mathrm C2: \mathrm{tr}(\mathrm {\bf C}_{\vec {xx}}) \leq P. 
\end{aligned} \label{sdp}
\end{equation}
Note that this is a  semidefinite programming (SDP) problem and it can be easily solved by
using numerical convex program solvers such as CVX \cite{cvx}. 

\section{Results and discussion} \label{results}

\begin{figure}[t]
	\centering {\includegraphics[trim = 6mm 7.1mm 0mm 8mm, clip, scale=0.15]{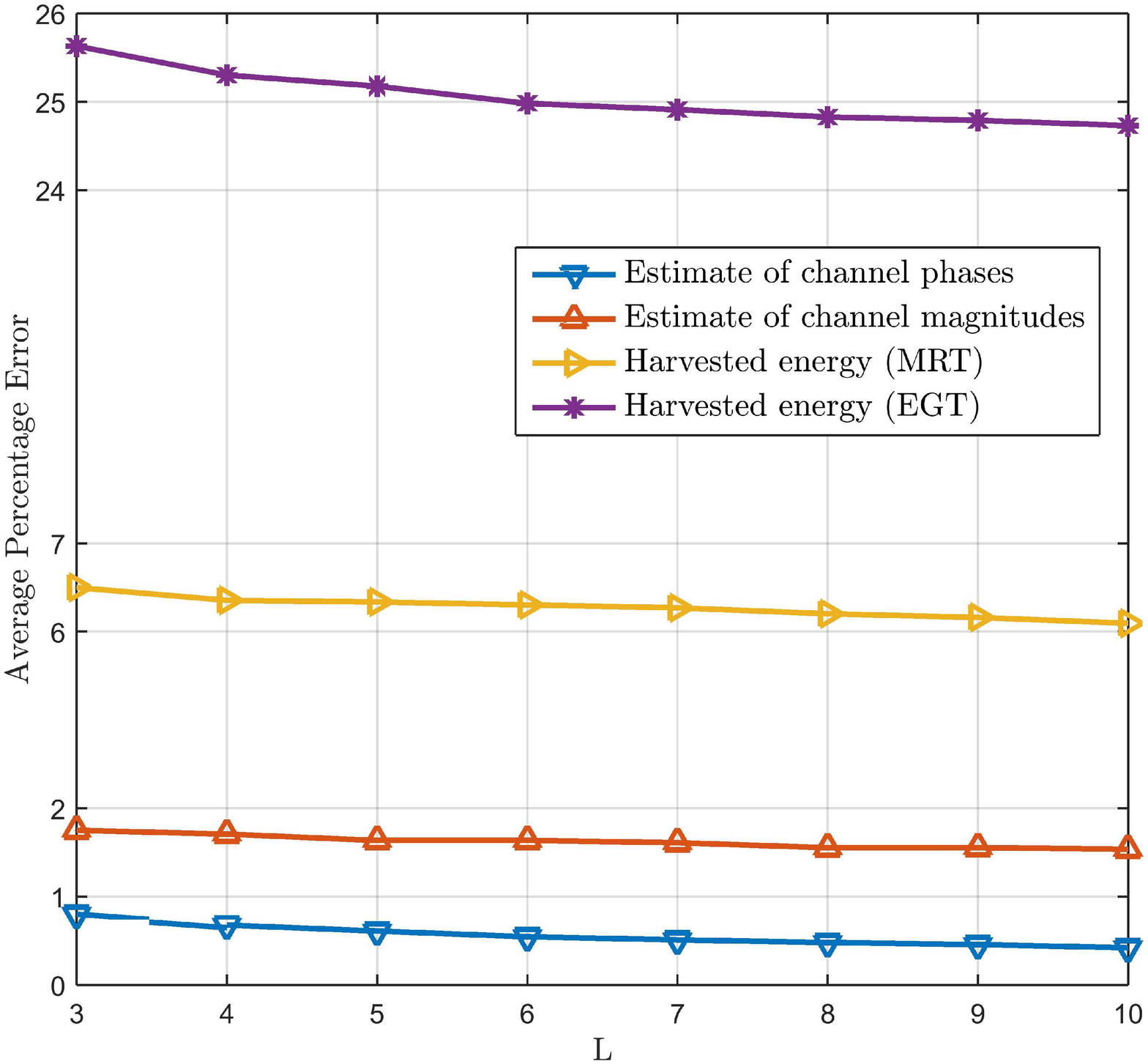}} 
	\caption{The behavior of the error percentage, for $K=4$ and SNR=$ 20$dB.}	 
	\label{csi} \vspace{-0.2cm}
\end{figure}

\begin{figure}[t]
	\centering {\includegraphics[trim = 6mm 7.5mm 0mm 8mm, clip, scale=0.15]{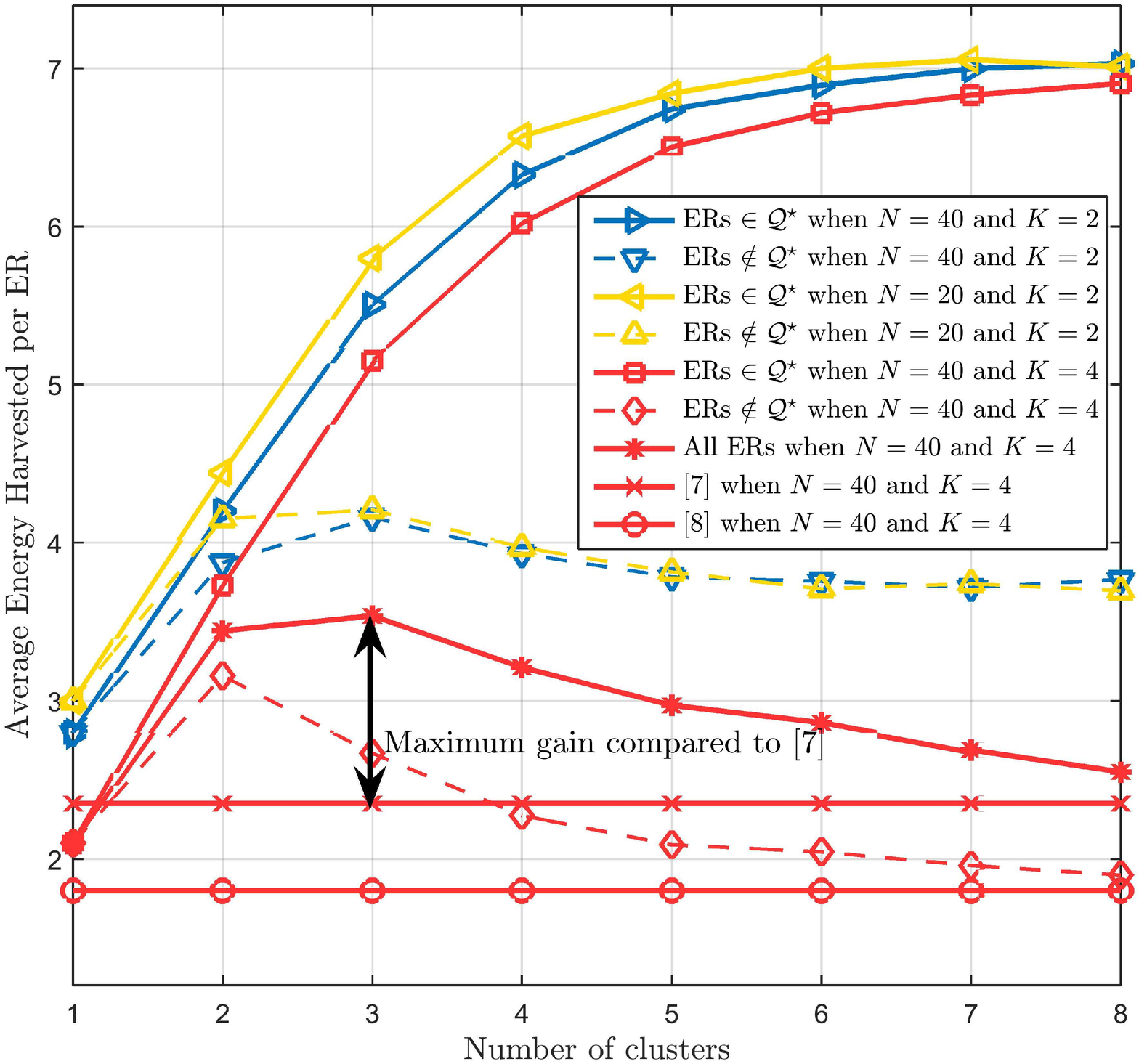}} 
	\caption{The behaviour of the harvested energy.}	 
	\label{opp} \vspace{-0.5cm}
\end{figure}

In this section, we present some simulation results and numerical evaluations to highlight the performance gains that can be achieved from our proposed schemes. In both simulations, the random channel amplitudes are uniform between $0.1$ and $1$, channel phase values are uniform between $0$ and $2 \pi$, and averaging is done over $1000$ iterations. We will start with the channel estimation. To this end, Fig. \ref{csi} illustrates how the the average error percentage changes with the amount of feedback, focusing on one ER. We can see that the phase estimation error and the magnitude estimation error is very low. The two graphs on harvested energy represent the average loss in harvested energy due to opting for RSSI based channel estimation, relative to the WET that can be achieved with MRT beamforming with perfect CSI. We can see that the loss is rather acceptable given the practicality of the proposed method compared to having perfect CSI at the ET. When comparing the two graphs, we can see that there is a significant improvement of going for MRT beamforming using the channel estimation techniques in this paper, compared to the EGT beamforming used in \cite{our_rssi}. For the selected parameters in this simulation, the improvement is approximately $20\%$.    

Fig. \ref{opp} illustrates the behavior of the average energy harvested per ER with the number of clusters. Note that the number of clusters being equal to one is equivalent to having no clustering, $\ie$, we try to maximize the minimum harvested energy among all $ N $ ERs. It is not hard to see that clustering is certainly useful.  For example, when $Q=3$, $N=40$ and $K=4$, we get an approximately $75\%$ improvement in the average energy harvested per ER due to clustering. It is rather obvious that ERs in $\mathcal{Q}^\star$ should harvest more energy, but for a given $Q$, the energy harvested by the ERs in $\mathcal{Q}^\star$ have decreased with both $N$ (due to having a lesser number of ERs in the selected cluster percentage wise) and $K$ (due to the beam being more directive). It is also interesting to note that the proposed opportunistic scheduling policy outperforms both \cite{oppor1} and \cite{oppor2} for the selected parameters. 
Due to space limitations, further interesting numerical evaluations and insights will be presented in future extensions of this work. \vspace{-0.2cm}

\section{Conclusions} \label{conclusions}

This paper has proposed a novel channel estimation methodology, and an opportunistic scheduling policy to be used in
a WET system consisting of multiple ERs. In the training stage, the ET transmits using a set of predefined codebooks, and each ER feeds back corresponding RSSI values to the ET. These values have been used for channel estimation. Based on the channel
estimates, the ERs have been grouped into clusters, and the most dense cluster has been selected. 
The beamformer that maximizes
the minimum harvested energy among all ERs in the selected
cluster have been found by solving a convex optimization problem. This beamformer is used to transfer power to the ERs using MRT beamforming, while achieving fairness over time.  \vspace{-0.1cm}

\footnotesize {\bibliography{bibfile}}

\end{document}